\documentclass{acm_proc_article-sp}
\usepackage{times}
\usepackage{helvet}
\usepackage{courier}
\usepackage{amssymb}
\usepackage{amsmath}
\usepackage{algorithmic}
\usepackage{algorithm}
\usepackage{subfigure}
\usepackage{graphicx}
\usepackage{multirow}

\begin{document}

\title{Gender Prediction in Social Media}
\author{James Smith\\
Computer Science and Engineering\\ Ohio State University\\
James.Smith@osu.edu
}

\maketitle

\begin{abstract}
In this paper, we explore the task of gender classification using limited network data with an application to Fotolog. We take a heuristic approach to automating gender inference based on username, followers and network structure. We test our approach on a subset of 100,000 nodes and analyze our results to find that there is a lot of value in these limited information and that there is great promise in further pursuing this approach to classification. 
\end{abstract}

\section{Introduction}
As social media becomes more and more integrated into our environment, it plays a bigger role in our daily lives. Because of this inclination, social media sites have become a major research area for computer scientists. The topic is becoming even more worthwhile as economic and business opportunities have sprouted up around these sites. These companies often have foundations built around the large amounts of user information that is readily available through social media sites. Social media sites also use this user information to automate things like profile customization, advertisement targeting, and interest prediction as well. However, the information that is given by users is often segmented and incomplete. Users may provide information such as name and age but could leave out other information such as race, interests or gender~\cite{zheleva2009join,abbasi2012real}. Thus, it has become increasingly important to predict other pieces of a user's information based on the subset of information that they provide~\cite{krosnick2005measurement,abbasi2013measuring}. Predicting users preferences and demographic information has a long history some of which can be found at ~\cite{kuma-etal11,SMM2014}.
Gender seems like one of the most basic pieces of information but can open the door to many other applications of information prediction. Given a user's actual name, it becomes almost trivial to predict their gender since we can simply evaluate whether it is a boy's name or a girl's name with the only discrepancies being if the name is not gender-specific (ex. Taylor, Sam, etc). However, sometimes we do not have access to a user's actual name and thus must use other information to predict their gender such as username (or screen name) and followers (or friends).

In this paper, we explore a heuristic approach to gender classification of users on the social media site Fotolog. Fotolog is a photo-blogging site in which users express themselves through online photo diaries. Currently, there are over 22 million members in over 200 countries using Fotolog. This site allows users to connect with each other by following user's photo posts. We explore to see how much information we can extract from a Fotolog user's screen name and come up with an algorithm that proves to be a fairly strong predictor of gender based on this information. 

\section{Literature Review}
There are considerable amount of work which show the possibility and effectiveness of using social media data to infer users' missing attributes \cite{murray2000inferring,li2012towards,murray2000inferring,marcus2006personality,abbasi2012lessons,conover2011predicting,mislove2010you}.
Rao et. al ~\cite{rao2010classifying} explored the classification of user attributes such as age, gender and political affiliation based on Twitter user language or other highly informal information. This was one of the first studies done on classification in the social media domain and built upon similar research that had been done in other domains such as classification based on email language, search queries, and blogs. The authors explored their dataset to find differences in follower-followee ratio, follower frequency, following frequency, response frequency, retweet frequency and tweet frequency. These differences were used to infer network structure. Interestingly, contrary to intuition, they found no correlation in the differences of these features and their relation to gender. In this study, the authors were able to achieve an accuracy of up to 72.33\% for gender prediction and found correlations between gender and use of emoticons, elipses, repetitive characters, repetitive exclamation, and the acronym ``OMG''.
Thus, based on these findings, we chose not to pursue gender prediction based on the follower sets for the users. This is because we wanted to make our solution as generalizable as possible and although we might find some sort of a correlation between the features in Fotolog, it would not be the case for all social media sites in general.  

Pennacchiotti and Popescu \cite{pennacchiotti2011machine} described a generalized machine-learning framework for user classification that relies on user profile, user tweeting behavior, linguistic content of user messages, and user social network features. The authors looked at predicting political affiliation, ethnicity, and whether or not the individual was a Starbucks fan. They were able to achieve an impressive 87.8\% accuracy on certain fields. This paper incited our interest in taking a heuristic approach to gender classification. They used certain trending topics and phrases for some of their classification that we decided would be interesting to apply to our data.

\section{Problem Statement}
The problem that we are trying to address is gender classification and prediction through very limited amounts of information. More specifically, given a list of screen names and followers, can we predict the gender of a user with a high degree of accuracy? We explore the application of this problem to the social media site Fotolog\footnote{http://us.fotolog.com}. We are given a set of 100,000 usernames and their followers as well as their genders. Based on this set, we must create an algorithm that can predict the genders of other users in the population.

\section{Solution}
To solve this problem, we tried several different approaches. The first problem that we ran into was that only 84,589 of the 100,000 nodes had known gender values. Thus, we had to throw out the remaining 15,411 nodes whose genders were unknown. This may have had a slight effect on our results as we were supposed to calculate our algorithm's error rate on 100,000 nodes. 

We first began by trying to find a list of trending topics and terms that are traditionally associated with women's interests (Oprah, style, etc.), however, we were unable to find a pre-defined set so we attempted to create our own. We essentially just brainstormed as many topics as we could think of that are traditionally associated with females and created a list out of that. We then parsed through the list of screen names and searched within each username to see if any of the topics appeared as a substring. If they did, we marked the user as female. Otherwise, we created a random number generator and assigned the user as female with a 70\% probability and male with a 30\% probability. These probabilities were assigned after we did research to find estimated overall population statistics for Fotolog users. We also experimented around with these probabilities and found that they were the most accurate yield. We tried using a 50/50 approach as well and found that it dropped accuracy to 60\%. This overall heuristic method produced very low accuracy (around 40\%) and thus was worse than simply guessing so we discarded it.

Our next approach was to use a list of male and female names that were publicly available online. We created separate lists for each set of names. We then used the same approach and searched to see if the screen name had a female name or male name as a substring and labeled them accordingly. We also used the same random guessing for the remaining user names. This produced an accuracy of 55.46\% for the set of data. We were not satisfied with this percentage.

We finally tried using just the set of female names since we wanted to see if it was more likely that females included their real names in their screen names and that males didn't necessarily follow that trend. We labeled the remaining data with the same methodology as before. This produced an accuracy of 63.95\%, which was much higher than the previous attempts. We then tried to slightly increase this percentage by adding a couple terms such as ``girl'' and ``love'' to the list of search words (these were not female names but words commonly associated with female screen names). These additions produced our final accuracy of 64.70\% for our data.

\section{Evaluation}
The heuristic approach seemed to work fairly well for the data that we had. It shows that there is a correlation between females and including female names in their screen names. 
We also found that some of the names that appeared in the list of female names were substrings of male names and that this might have had an effect on the prediction rate as well. For instance, Bert is considered a girl's name but is contained in the name Robert that is a boy's name. This might also have been a cause for the percentage being lower when we searched boys' and girls' names as opposed to just girls' since we first searched through the girl names then the boy names. This problem could be fixed in a future implementation by checking which occurs in a longer substring of the screen name. 

We believe that we can greatly increase the accuracy by including a more robust list of female terms as well as searching through the screen names for certain characteristics such as the use of emoticons, elipses, repetitive characters, repetitive exclamation, and the acronym ``OMG''. In future work, we would also eliminate the 70\% rate of prediction for gender being female since the population parameters are often unknown. This would create a more generalizable solution for social media sites that have little to no information published about their user-base. However, it would be fair to assume knowledge of estimated population gender parameters for the larger social media sites such as Twitter or Facebook.

\section{Conclusion}
User classification and gender prediction are very important topics in the growing research field of social networks. They provide interesting opportunities in the fields of advertisement, profile customization, and interest prediction that could pave the way to future innovations. 
In this paper, we proposed a simple heuristic approach to gender prediction that yields an accuracy of 64.70\% for the data set that we had. This is fairly low relative to other similar work that has been done in the area but operates on a very small set of starting information. Because of this constraint, it makes the algorithm fairly generalizable to most other social media sites however makes it difficult to get very high rates of accuracy. There is still much more work that needs to be done on this algorithm and in this area. It still needs to be tested on sets of data from other social media sites to see if the trend that was found only applies to Fotolog or if it can be generalized to the social media world. This algorithm also operates based on screen names which is a field that most social networking sites are starting to stray away from and move towards users using their actual names. The actual name would great increase the accuracy of this algorithm and allow for much higher rates of success. 

A heuristic approach to gender prediction seems to have a lot of merit but is a pretty daunting task in that no standards have been created for gender associated terms in the social sphere and thus there is very little prior research banks to build on top of. However, there does seem to be a fairly strong correlation between gender and the user of gender-specific names in a user's screen name. we believe that the accuracy of this methodology could be greatly increased through the accumulation and integration of gender-specific terms and popular culture (such as love, Justin Bieber, etc) and the use of pattern recognition in screen names to search for features such as recurring letters and capitalization. All of these methodologies come with a computational trade-off in that they increase the amount of time required for the algorithm to run exponentially.

This heuristic method could also be augmented with a machine-learning algorithm that would allow the program to automatically detect trends in terms used in screen names that correlate to gender. Those terms could also be added to the heuristics word bank. This would address the issue of applying this algorithm to sites that contain foreign content or English slang so that we can attribute for non-name and non-topic based terms used within screen names.

\newpage 
\bibliographystyle{abbrv}
\bibliography{citations}

\end{document}